\documentclass[letterpaper]{article} 
\usepackage{aaai2026}
\usepackage{multirow}
\usepackage{times}  
\usepackage{helvet}  
\usepackage{courier}  
\usepackage[hyphens]{url}  
\usepackage{graphicx} 
\usepackage{natbib}  
\usepackage{caption} 
\usepackage{algorithm}
\usepackage{algorithmic}

\usepackage{newfloat}
\usepackage{listings}
\DeclareCaptionStyle{ruled}{labelfont=normalfont,labelsep=colon,strut=off} 
\floatstyle{ruled}
\newfloat{listing}{tb}{lst}{}
\floatname{listing}{Listing}
\usepackage[utf8]{inputenc} 
\usepackage[T1]{fontenc}    
\usepackage{url}            
\usepackage{booktabs}       
\usepackage{amsfonts}       
\usepackage{nicefrac}       
\usepackage{microtype}      
\usepackage{xcolor}         
\usepackage{graphicx}
\usepackage{lipsum} 
\usepackage{listings}
\usepackage{xcolor}  
\usepackage{amsmath} 
\usepackage{pifont}
\newcommand{\xmark}{\ding{55}}

\lstdefinestyle{mypython}{
    language=Python,
    basicstyle=\ttfamily\small,
    keywordstyle=\color{blue},
    commentstyle=\color{gray},
    stringstyle=\color{red},
    numbers=left,
    numberstyle=\tiny\color{gray},
    stepnumber=1,
    numbersep=5pt,
    backgroundcolor=\color{white},
    showspaces=false,
    showstringspaces=false,
    showtabs=false,
    frame=single,
    tabsize=2,
    captionpos=b,
    breaklines=true,
    breakatwhitespace=false
}

\title{Proc3D: Procedural 3D Generation and Parametric Editing of 3D Shapes with Large Language Models}
\author{
Fadlullah Raji\textsuperscript{\rm 1},
Stefano Petrangeli\textsuperscript{\rm 2},
Matheus Gadelha\textsuperscript{\rm 2},
Yu Shen\textsuperscript{\rm 2},
Uttaran Bhattacharya\textsuperscript{\rm 2},
Gang Wu\textsuperscript{\rm 2}
}

\affiliations{
\textsuperscript{\rm 1}University of South Florida\\
\textsuperscript{\rm 2}Adobe Research\\
fraji@usf.edu, \{gadelha, shenyu, petrange, ubhattac, gawu\}@adobe.com
}

\usepackage{bibentry}

\begin{document}

\vspace{-3mm}
\twocolumn[{%
\renewcommand\twocolumn[1][]{#1}%
\maketitle
\begin{center}
    \centering
    \captionsetup{type=figure}
    \includegraphics[width=0.95\textwidth, height=0.4\textheight, keepaspectratio]{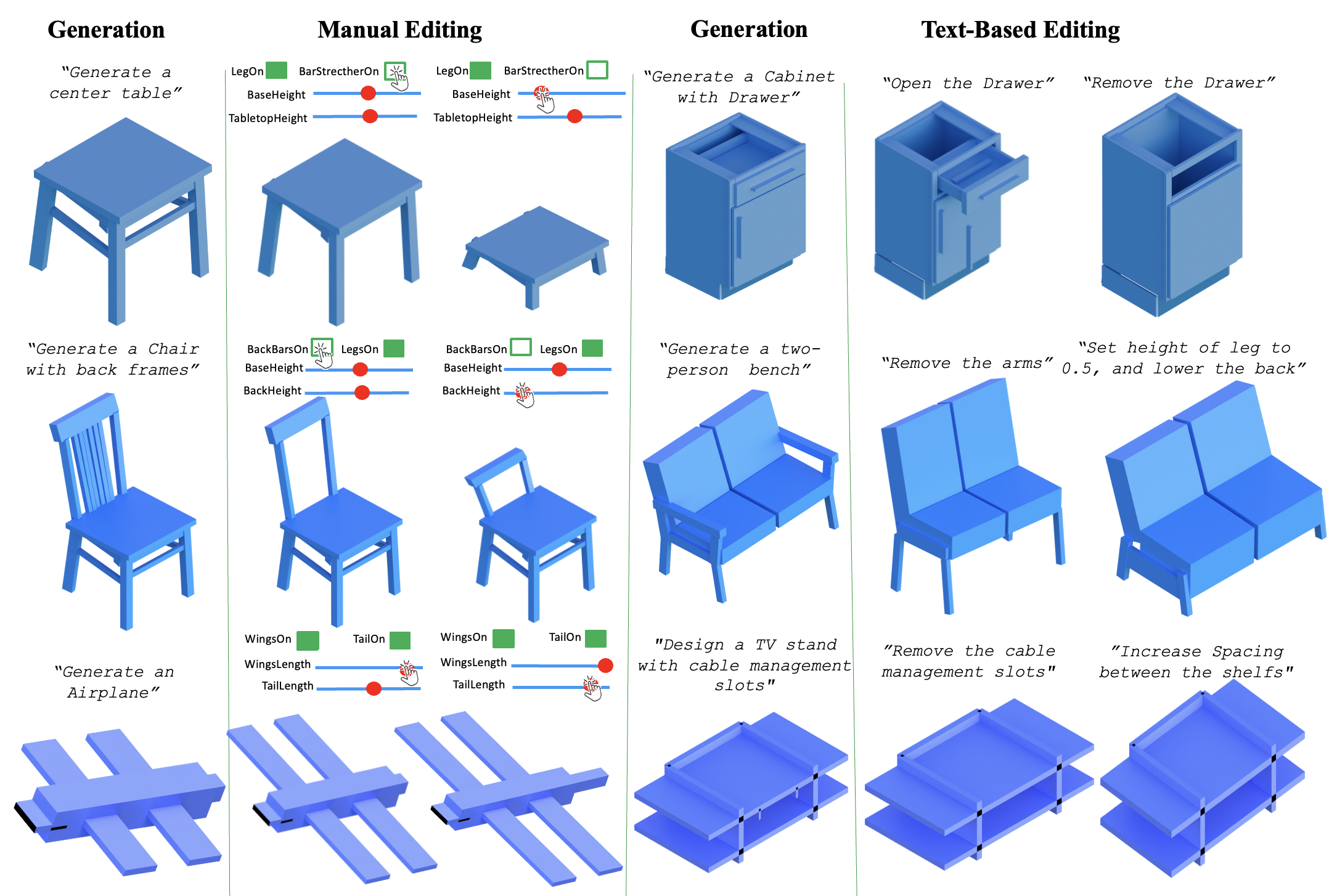}
        \caption{\textbf{Qualitative results of Proc3D.} Proc3D facilitates the generation of editable 3D models, and relative parameters from Large Language Models (LLMs) using Procedural Compact Graph (PCG) representation, which provides an efficient, compact, and editable graph-based structure for 3D models. These graphs can be directly modified in real time, using intuitive manual controls, such as sliders and checkboxes. Additionally, Proc3D enables text-based editing of the PCG with the assistance of the LLM, allowing users to make modifications through concise or detailed natural language instructions.}
        \label{fig:results}
\end{center}%
}]


\begin{abstract}
Generating 3D models has traditionally been a complex task requiring specialized expertise. While recent advances in generative AI have sought to automate this process, existing methods produce non-editable representation, such as meshes or point clouds, limiting their adaptability for iterative design. In this paper, we introduce Proc3D, a system designed to generate editable 3D models while enabling real-time modifications. At its core, Proc3D introduces procedural compact graph (PCG), a graph representation of 3D models, that encodes the algorithmic rules and structures necessary for generating the model. This representation exposes key parameters, allowing intuitive manual adjustments via sliders and checkboxes, as well as real-time, automated modifications through natural language prompts using Large Language Models (LLMs). We demonstrate Proc3D’s capabilities using two generative approaches: GPT-4o with in-context learning (ICL) and a fine-tuned LLAMA-3 model.
Experimental results show that Proc3D  outperforms existing methods in editing efficiency, achieving more than \textbf{400× speedup} over conventional approaches that require full regeneration for each modification. Additionally, Proc3D improves ULIP scores by 28\%, a metric that evaluates the alignment between generated 3D models and text prompts. By enabling text-aligned 3D model generation along with precise, real-time parametric edits, Proc3D facilitates highly accurate text-based image editing applications.
\end{abstract}
\section{Introduction}
\label{sec:intro}
High-quality 3D modeling remains a labor-intensive task requiring specialized expertise. Recent text-driven generative AI can produce point clouds \cite{pointe2022, Text2PointCloud2023}, meshes \cite{TextMesh2024, Text2Mesh2022, siddiqui2023meshgpt}, implicit fields \cite{Shape2023, li2023diffusionsdf, cheng2023sdfusion}, and other representations \cite{poole2023dreamfusion, lu2024direct2, lin2023magic3d, chen2023fantasia3d}. However, these formats are inherently static: point clouds lack explicit connectivity for targeted edits; meshes demand costly re-meshing or constraint solving for even minor changes \cite{Fruhauf_2024_CVPR, achlioptas2023shapetalk}; and implicit or neural fields do not expose human-interpretable parameters, making iterative refinement and localized adjustments cumbersome.
Procedural 3D modeling, where shapes are defined by parametric operations and algorithmic rules offers dynamic control and precise editability \cite{geocode, greff2021kubric, he2021semi, procthor2022, infinigen2023infinite, 3DGPT, scenex}. Yet crafting procedural programs demands deep domain knowledge, and existing AI-driven approaches focus on driving predefined procedural templates \cite{geocode, 3DGPT, scenex, stekovic2024pytorchgeonodes}, constraining user creativity to limited parameter ranges.

    In this work, we present \textbf{Proc3D}, a unified system for generating and editing procedural 3D graphs directly from text. At its core is the \emph{Procedural Compact Graph (PCG)}, a language-native, engine-agnostic graph representation that encodes 3D objects as a compact sequence of high-level operations, exposes editable parameters on each node for true real-time, localized edits via GUI controls or natural-language follow-ups, and compiles seamlessly to multiple runtimes without API-specific dependencies. We collect a mesh-derived PCG–instruction pairs and fine-tune a LLaMA-3 model for robust PCG generation. Our evaluations show that Proc3D achieves 4–10× more compact representations than code-based DSLs (Table~\ref{tab:representation_comparison}); improves ULIP alignment by 28\% over text-to-3D baselines (Table~\ref{tab:ULIP_comparison}); enables local edits 400× faster than full-regeneration methods, and maintains strong in-distribution and out-of-distribution performance with detailed ablations on LLM choices and interpreter latency.

\textbf{Our contributions are as follows:}
\begin{itemize}
\item  We develop Proc3D, a system that enables 3D generation from high-level text input and intuitive edit of 3D models through a compact graph representation.

\item  We introduce a compact and efficient representation of procedural 3D graphs that simplifies their inherent complexity, termed \textit{procedural compact graph (PCG)}.

\item  To enable robust training and evaluation, we curate a large dataset of procedural 3D graphs along with corresponding instructions, and fine-tune a trained LLaMA-3 model for robust generation of 3D models.
\end{itemize}

\begin{figure*}[t]
\vspace{-3mm}
\centering
\includegraphics[width=0.96\textwidth]{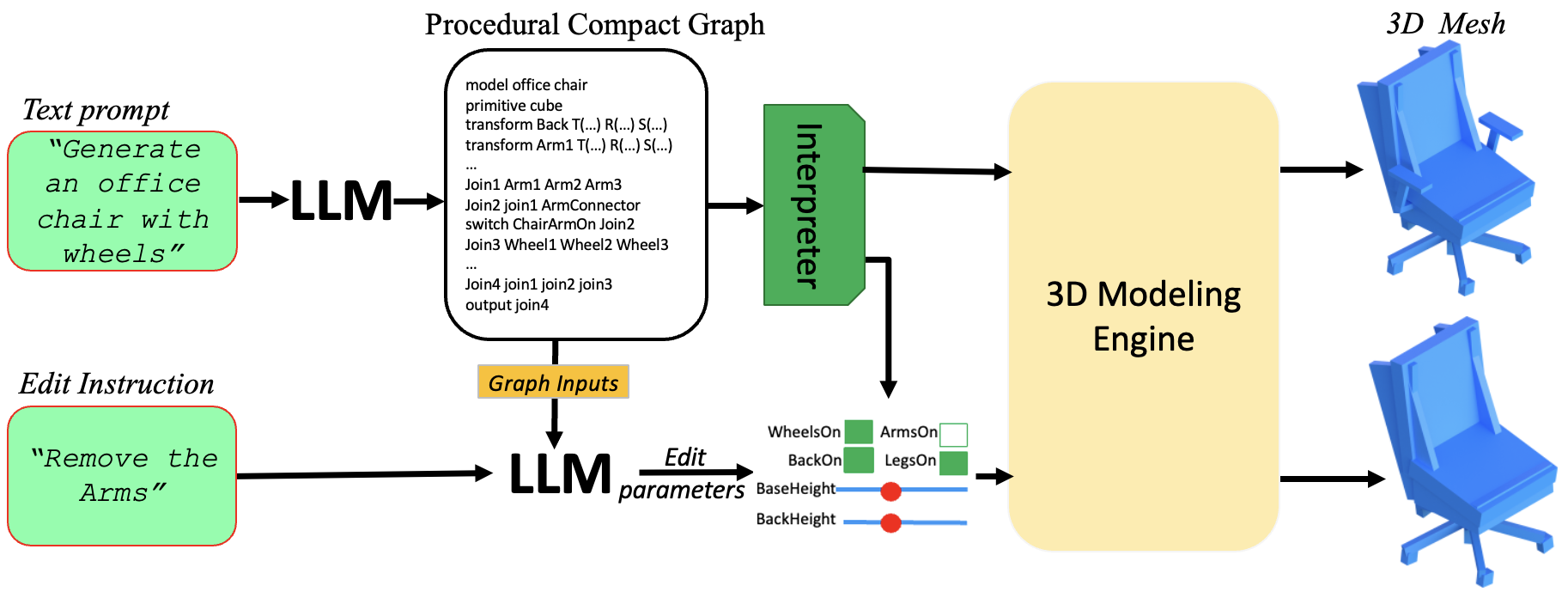} 
\caption{\textbf{3D Generation and Editing Pipeline}. Given a text prompt, a Large Language Model (LLM) constructs the Procedural Compact Graph (PCG), which is interpreted into software-specific formats suitable for 3D modeling engine such as Blender or Unity3D. Subsequent edits submitted via text prompt are integrated by the LLM to update the graph and, consequently, the 3D mesh.}
\label{fig:3D generation and edit pipeline}
\vspace{-6mm}
\end{figure*}

\section{Related Work}
\label{sec:related_work}

\paragraph{Text to 3D Shape Generation.}
Text-to-3D generation has seen significant advancements, with various methods improving different aspects of the process. Early approaches like CLIP-Mesh \cite{khalid2022clipmesh} utilized CLIP \cite{clip} model for zero-shot 3D synthesis. Since then, several works have rapidly advanced the generation of 3D objects using meshes \cite{TextMesh2024, Text2Mesh2022, siddiqui2023meshgpt} and signed distance fields (SDF) \cite{cheng2023sdfusion, li2023diffusionsdf}. Text-guided methods have also evolved rapidly, particularly through score distillation sampling (SDS) \cite{poole2023dreamfusion, lin2023magic3d, chen2023fantasia3d, tang2023dreamgaussian, wang2023prolificdreamer}, which distills pre-trained text-to-image diffusion models to improve quality and text-to-3D alignment. Despite these advancements, the representations through which these 3D models are generated, makes control and editing difficult. Proc3D introduces Procedural Compact Graph (PCG), a language-native graph of high-level operations that LLMs can generate and parse. Unlike static mesh or implicit-field methods, PCG inherently exposes node-level parameters for targeted, real-time edits without requiring full regeneration.

\paragraph{Procedural 3D Modeling.}
Procedural modeling has long been pivotal for generating structures like plant models \cite{Prusinkiewicz1996, Gasch2022} and urban environments \cite{Parish2001, Muller2006}, often via procedural graphs which may also be used to generate natural scenes \cite{Gasch2022, Zhang2019} and cityscapes \cite{Lipp2011, Yang2013, Vanegas2012, Talton2011}. Tools like Blender offer drag-and-drop graph construction, but extensive parameter tuning remains labor-intensive. Recent generative AI approaches help automate some aspects: ShapeAssembly \cite{jones2020shapeAssembly} uses code-to-code interpolation, Geocode \cite{geocode} maps sketches or point clouds to parametric graphs, and Infinigen \cite{infinigen2023infinite} generates infinite landscapes and indoor scenes \cite{infinigenindoor}. Large language models have also been used to reduce manual editing \cite{scenex, 3DGPT}, but procedural graph creation itself still relies heavily on manual creation, limiting their broader automation. Our work is the first to enable direct generation of procedural 3D graphs from text instructions, opening new avenues for 3D modeling.
\paragraph{Large Language Models for 3D Generation.}

Large language models (LLMs) encode vast knowledge into language-based representations, enabling broad applicability across tasks~\cite{Raffel2020, Bubeck2023, Chowdhery2022, openai2024gpt4, BERT}. Initially developed for text, LLMs now extend to mathematical reasoning~\cite{Wei2022, imani2023}, planning~\cite{Song2023, Huang2022, Zhao2023}, and code generation~\cite{codex, codellama}, motivating their adaptation for 3D content creation. However, 3D generation remains fundamentally difficult due to the spatial, hierarchical, and topological complexity of geometric data. Recent works have explored mesh-based and procedural approaches, but limitations remain. Autoregressive models like MeshXL~\cite{meshxl} and LLaMA-Mesh~\cite{llamamesh} serialize geometry as vertex-face sequences, which are difficult to edit or interpret in natural language. Procedural methods such as 3D-GPT~\cite{3DGPT} and SceneCraft~\cite{scenecraft} rely on fixed templates or asset libraries with limited parameter control, constraining expressiveness. BlenderAlchemy~\cite{huang2024blenderalchemy} enables vision-language editing of 3D scenes but lacks support for generative modeling from scratch. Meanwhile, L3GO~\cite{L3goYamada2024} and 3D-Premise~\cite{3dpremise} generate Blender Python code via prompting, but are tightly coupled to Blender's engine and produce verbose, brittle scripts.

To overcome these issues, we introduce a simplified procedural modeling framework that reduces task complexity for LLMs by leveraging structured, interpretable graph representations. Our system, Proc3D, enables editable, engine-agnostic 3D generation directly from natural language prompts.

\section{Proc3D}
\label{sec:method}
\paragraph{System Overview.}
An overview of our system is shown in Figure \ref{fig:3D generation and edit pipeline}. The system facilitates both generation and intuitive editing of procedural 3D models using natural language inputs, by leveraging LLMs and a compact representation for procedural 3D graphs. At the core of the system is  \textit{Procedural Compact Graph (PCG)}, a simplified language designed to reduce the complexity of traditional procedural modeling, and making it feasible for LLMs to efficiently generate and interpret 3D models.

To generate 3D models, users provide natural language instructions, which are then translated into the corresponding procedural representation. This representation is decoded into input parameters, such as booleans to toggle parts or continuous values to adjust dimensions. A software-specific interpreter converts the graph into executable code—such as Python scripts for Blender—where the 3D model is rendered. Users can perform parametric edits using sliders or checkboxes linked to the input parameters, or automated edit via additional natural language commands (e.g., "\texttt{Make the legs taller}"), enabling real-time modifications.

In the following subsections, we detail our contributions: the design of the \textit{Procedural Compact Graph (PCG)}, the translation of 3D meshes into PCG for large dataset curation, and the process of generating and editing 3D models via natural language inputs.
\subsection{Procedural Compact Graph (PCG)}
\label{sec:pcg}
In traditional 3D modeling software like Blender, users create procedural 3D models by connecting geometric nodes through a drag-and-drop user interface (UI). While intuitive, these node-based graphs often become complex, especially in platform-specific code formats, creating challenges for LLMs due to context length limitations and syntax errors \cite{li2024longcontext}.
\begin{figure}  
    \centering
    \includegraphics[width=\linewidth]{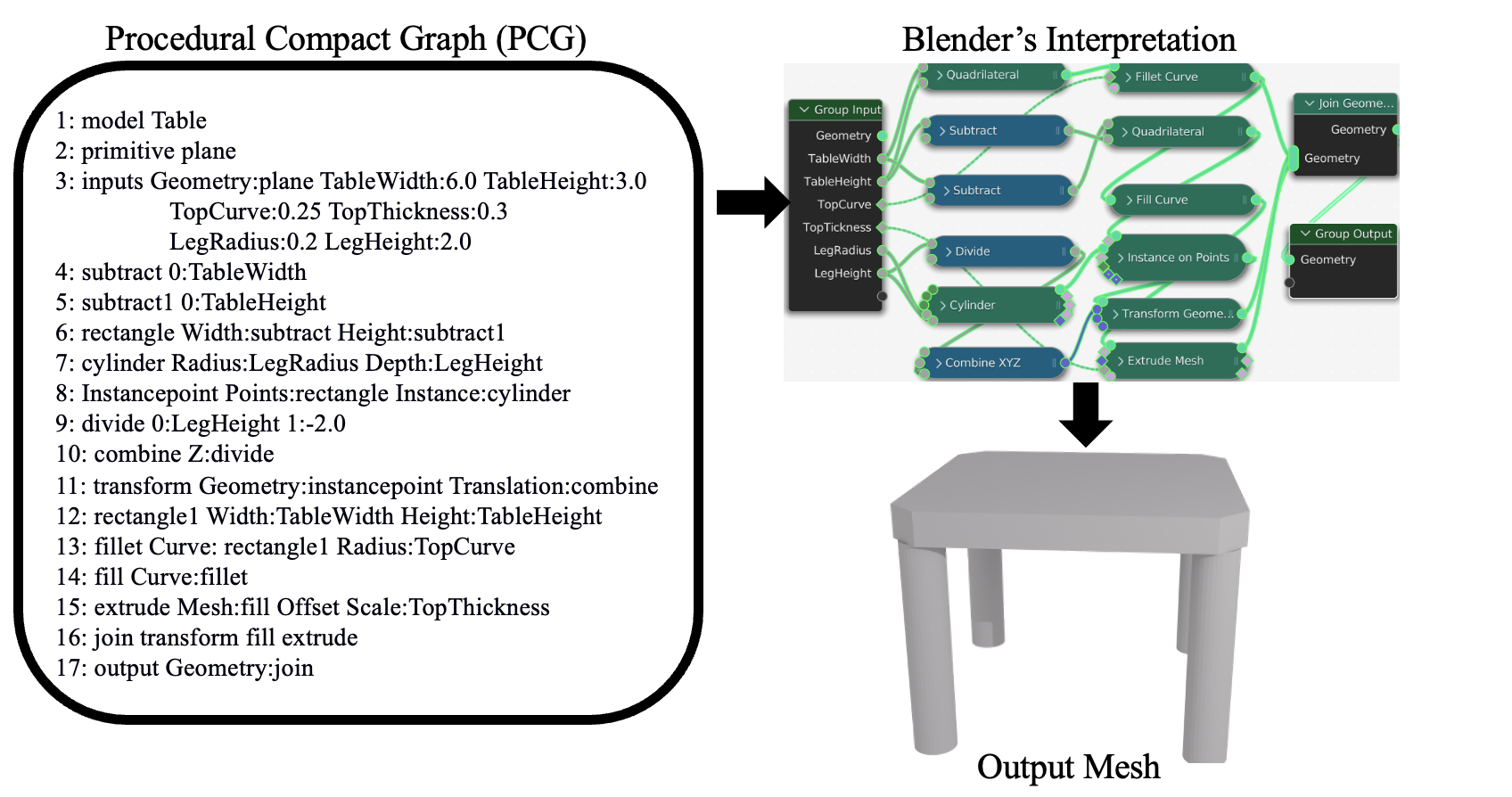}
    \caption{\textbf{An example of the procedural compact graph (PCG)}. Each line is a node in the procedural 3D graph that defines the operation performed on the connected node. The graph is interpreted into blender's geometric node where the mesh can be visualized.}
    \label{fig:pcg_example}
    \vspace{-3mm}
\end{figure}
To address this, we propose the Procedural Compact Graph (PCG), a simplified representation that abstracts procedural operations into a concise format, reducing context overhead by 4-10X compared to platform-specific codes. For example, the PCG example shown in Figure \ref{fig:pcg_example} achieves a 5X reduction over Infinigen's code \cite{infinigen2023infinite} (see supplementary material) and over 10X compared to Blender's Python code for the same 3D mesh. Each PCG line represents a node or operation with defined parameters and connections, simplifying 3D model generation and interpretation by LLMs. Detailed node definitions are provided in the supplemental material. Next, we describe how to extract this representation from hierarchically segmented 3D meshes.

\begin{figure}
    \centering
    \includegraphics[width=\linewidth]{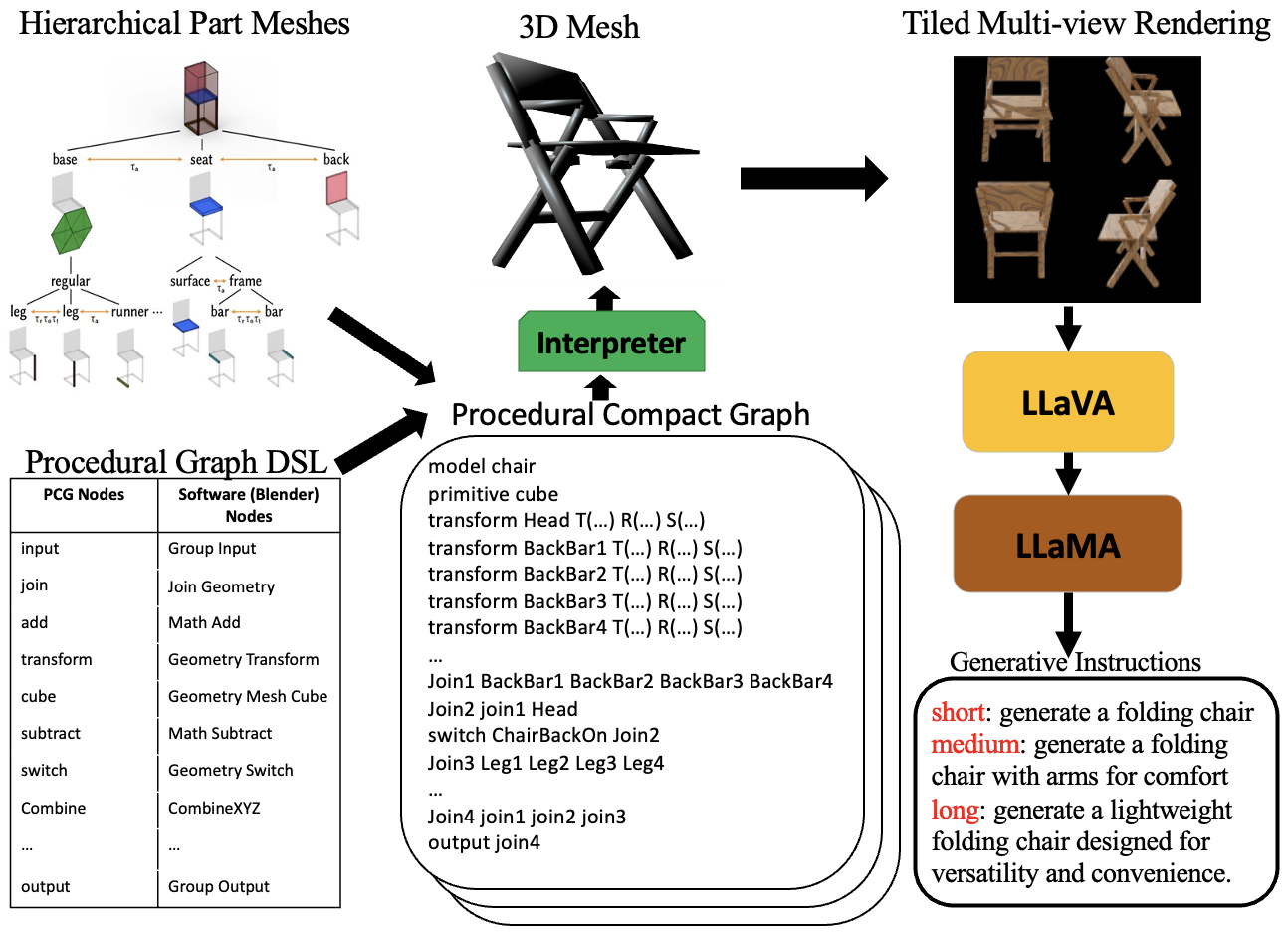}
    \caption{\textbf{Data Collection Pipeline}. Given a dataset of the hierarchical part meshes\cite{mo2019partnet, mo2019structurenet}, we extract the PCG representations, and rendered multi-view images of the 3D model. Then, using LLaVA, we collect detail caption of the rendered views, which is passed through LLaMA-3 70B for the generation of corresponding instructions.}
    \label{fig:data collection pipeline}
\end{figure}

\paragraph{Extracting PCG from 3D Meshes.}
\label{sec:extraction}
 Procedural 3D graphs effectively represent parametric 3D shapes but are complex to generate, typically requiring domain experts and creating a data collection bottleneck. To address this, we repurpose hierarchically segmented 3D meshes. Our pipeline begins with a large dataset of hierarchical 3D part models \cite{mo2019structurenet, mo2019partnet}, where shapes are assemblies of parts represented as nodes. These nodes are connected by edges that signify physical attachments and hierarchical parent-child relationships, indicating how larger parts comprise smaller sub-components. Leaf nodes, or atomic parts, are represented using cuboid geometries (see Figure \ref{fig:data collection pipeline}). Using the PartNet \cite{mo2019partnet} mesh dataset, we perform two key processes to extract PCG compositions, detailed below, and Psudo-code of the extraction process in Figure \ref{fig:psuedo_code} of the supplemental material.

A. \textbf{Recursive Combination of Parts.}
To construct a procedural 3D graph from semantically segmented meshes, we recursively traverse the hierarchy from root to leaf nodes. Each node represents a terminal component (e.g., chair back, table leg) and connects to its parent via a Join Node in PCG. Multiple leaf nodes with the same semantic name, like chair legs, are merged into a single child node before attaching to the parent. The Join Node integrates these components into a cohesive model, ensuring structural integrity. Transformations applied to a parent node, such as translation or scaling, automatically propagate to child nodes, allowing consistent modifications across the entire model.

B. \textbf{Exposing Editable Properties.}
At each recursive step, key properties such as rotation, scaling, and translation are exposed for precise adjustments to individual parts. Each child node includes a switch to enable or disable components, allowing seamless manipulation. For instance, toggling a chair's armrests \textit{on} or \textit{off} affects all sub-components under the \texttt{armrest} label, such as padding or supports. Major structural parts, like the chair base, are adjusted as a whole rather than individually modifying legs, bar stretchers, or wheels. Changes to the base's properties, such as height or scale, automatically propagate to related components, maintaining proportionality. This approach allows dynamic object modifications without regenerating the entire procedural graph.

\paragraph{ Instruction-Graph Data Collection.}
\label{sec:data collection}
Our data collection pipeline, shown in Figure \ref{fig:data collection pipeline}, converts hierarchical 3D meshes from the PartNet dataset \cite{mo2019partnet} into PCG representations, resulting in 21K models across five categories: Chairs, Tables, Trash Cans, Storage Furniture, and Beds. Multi-view images of the PCG models are then rendered from front-facing cameras to capture their full geometry and structure, ensuring detailed visualization from various angles. These multi-view images are compiled into tiled images, which are analyzed by LLaVA \cite{liu2023llava} to generate detailed text descriptions, capturing attributes like shape, geometry, and spatial relationships. The captions produced by LLaVA are then processed by LLaMA-3 to generate structured generative instructions tailored to three levels of detail: \textit{short}, providing concise summaries; \textit{medium}, offering detailed feature descriptions; and \textit{long}, covering comprehensive structural and contextual information. This process produces 63K instruction-graph pairs, enabling robust training and evaluation.
\begin{figure*}[t]
    \centering
\includegraphics[width=0.96\linewidth]{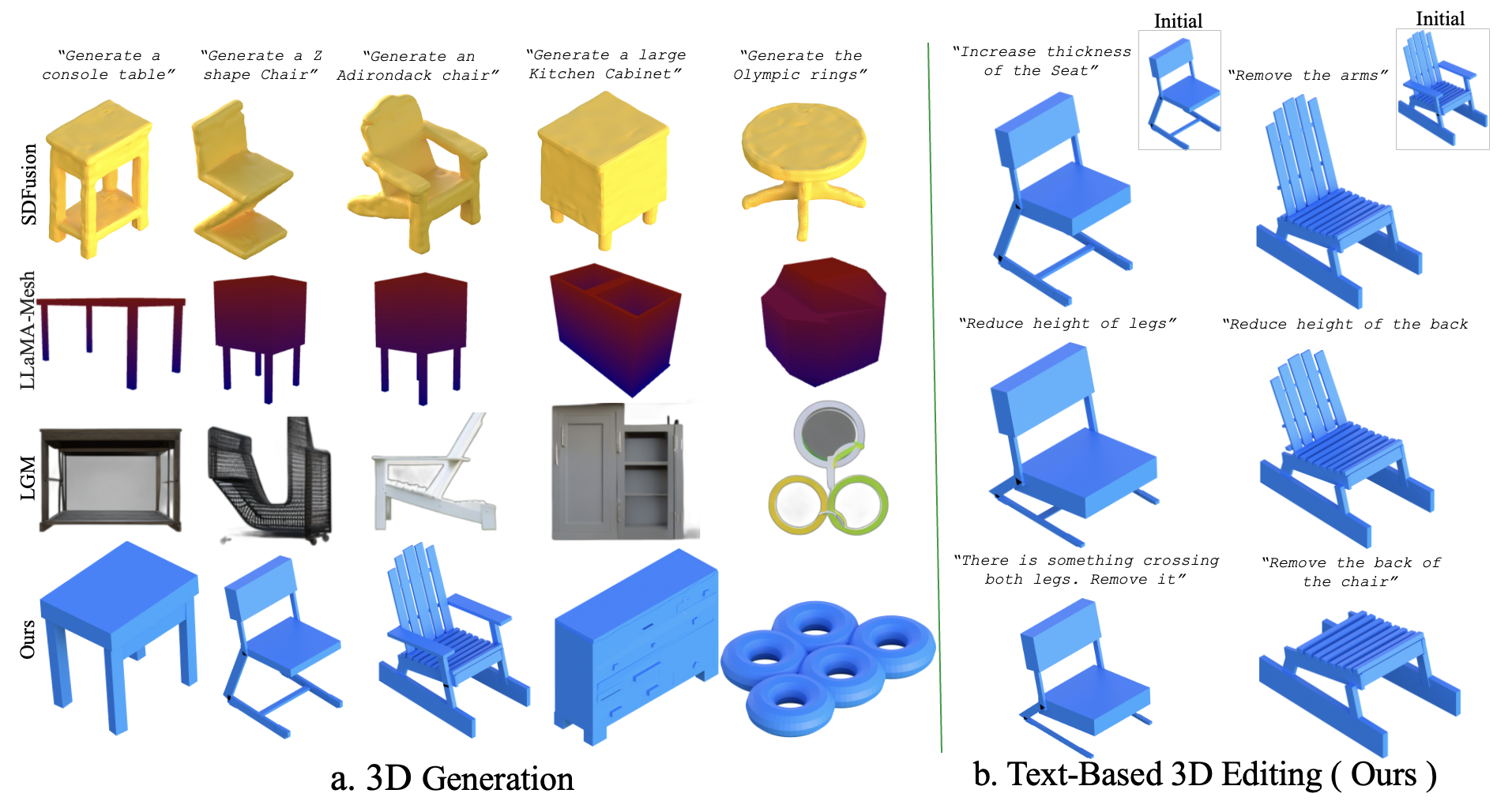}
    \caption{ \textbf{Comparative Analysis of 3D Object Generation and Editing.} (a) 3D Generation: Comparison of different methods for generating 3D models from text prompts, including SDFusion \cite{cheng2023sdfusion}, LLaMA-Mesh \cite{llamamesh}, LGM \cite{tang2024lgm}, and our proposed method (Ours). Our method achieves more consistent, structured outputs across various prompts. (b) Text-Based 3D Editing: Our system introduces representation that enabled targeted edits on the generated 3D objects from text prompts. Other methods do not support editing the generated 3D models.
    }
    \label{fig:qualitative_compare}

\end{figure*}
\subsection{3D Generation and Parametric Edits}
This subsection outlines our system's core functionalities, leveraging LLMs for text-based generation and parametric editing of 3D objects. Users can create 3D content and interactively modify object parameters via natural language. This interaction is facilitated by PCG, which serves as an intermediary representation of the object's geometric structure and connectivity. The graph-based approach ensures that generated objects remain parametric and adaptable throughout their lifecycle.

\paragraph{3D Object Generation.}
The generation process starts with a user-provided text prompt processed by the LLM. For example, a prompt like “\texttt{Generate an office chair with wheels}” is interpreted by the model to construct the corresponding PCG, which encodes the object's geometric and topological properties, including part connectivity, transformations, and relationships. This PCG is then passed to an interpreter that converts it into a software-specific format compatible with modeling environments like Substance 3D, Blender, or Unity. The interpreter maps the procedural definitions to commands understood by these platforms, making the graph's geometric instructions executable.
\paragraph{Parametric Edits and Interactive Modification.}
Our system supports robust parametric editing, enabling users to modify objects manually or via text-based instructions. Edits are applied at the PCG level, ensuring structural coherence and preserving the object's integrity. For example, a command like “Remove the arms” prompts the LLM to update relevant PCG parameters, such as removing components or adjusting transformation parameters. The graph-based structure ensures precise and efficient modifications, as changes to one component automatically update related transformations to maintain consistent geometry. The updated PCG is then recompiled into a software-specific representation for immediate visualization, enabling real-time iterative design through successive text inputs.



\section{Experiment}

We present experiments to evaluate the generation, and editing of procedural 3D models from our framework. 
For these experiments, we utilize both GPT-4o\cite{openai2024gpt4} and the LLaMA models (8B and 70B parameters) \cite{lamma3}.

\begin{table}
\centering
\vspace{-5mm}
{\small 
\setlength{\tabcolsep}{2.3pt}
\small 
\caption{\textbf{Comparative Performance of LLM-Generated Representations.} PCG representation demonstrates superior abstraction with the highest compile rate and the lowest generation time, highlighting significant advancements in efficiency and capability.}
\label{tab:representation_comparison}
\begin{tabular}{l p{1.2cm} p{1.4cm} p{1.2cm}}
\toprule
\textbf{Representation} & \textbf{Compile Rate (\%)} & \textbf{Avg. Token Length} & \textbf{Gen. Time (s)} \\ 
\midrule
Blender Geo. Node           & 0                         & 6048                        & 62                      \\
Infinigen\cite{infinigen2023infinite}         & 5                         & 3403                       & 50                      \\
Blender Code \cite{3dpremise}                   & 30                         & 2789                        &  43                     \\
LLaMA-Mesh \cite{llamamesh}                   & 45                         & 3189                        &  25                     \\
\textbf{PCG (Ours)    }                & \textbf{89}                         & \textbf{702}                        & \textbf{9}                     \\
\bottomrule
\end{tabular}}
\vspace{-4mm}
\end{table}

\paragraph{Datasets.} 
To enable few-shot ICL, we manually annotated 1,000 examples, where humans observed rendered 3D objects and provided generative instructions (e.g., "a table with a single central leg"). For the remaining graphs in the dataset, we applied the automated method outlined in \ref{sec:data collection}.
\paragraph{Evaluation Metrics.} To evaluate the effectiveness of our proposed system, we utilize three primary metrics: \emph{Compile Rate} to assess how often the generated 3D procedural graphs are syntactically correct and can be successfully compiled. \emph{Similarity Measure} to quantify the geometric similarities between the generated and the closest reference 3D models in the training data. Lastly, the \emph{ULIP} (Unified Language and Image Pre-training) \cite{xue2022ulip} score assesses the perceptual loss between the input text prompt and the generated shape, thus providing insight into the alignment between the 3D shape and the input prompts.

\vspace{-3mm}
\begin{table}[t]
\centering
\caption{\textbf{Comparison of ULIP Scores and Inference Time Across Methods.}
Our approach using finetuned LLaMA models (L-70B-F and L-8B-F) and non-finetuned GPT-4o achieves the highest ULIP score, outperforming SDFusion, AutoSDF, Shap-E, and LLaMA-Mesh.}
\label{tab:ULIP_comparison}
\scriptsize
\resizebox{\columnwidth}{!}{%
\begin{tabular}{ccccc}
\toprule
\textbf{Method} & \textbf{Editable} & \textbf{ULIP $\uparrow$} & \textbf{Gen. Time $\downarrow$} & \textbf{Edit Time $\downarrow$} \\
\midrule
SDFusion~\cite{cheng2023sdfusion} & \xmark & 0.11 & 8s  & 8s  \\ 
AutoSDF~\cite{autosdf2022} & \xmark & 0.07 & \textbf{4s} & 4s  \\
Shap-E~\cite{Shape2023} & \xmark & 0.10 & 30s  & 30s  \\
LLaMA-Mesh~\cite{llamamesh} & \xmark & 0.08 & 25s  & 25s  \\
\midrule
\textbf{Ours (L-70B-F)} & \textbf{\checkmark} & \textbf{0.15} & 47s  & \textbf{0.01s} \\
\textbf{Ours (L-8B-F)}  & \textbf{\checkmark} & 0.13 & 10s  & \textbf{0.01s} \\
\textbf{Ours (GPT-4o)}  & \textbf{\checkmark} & 0.14 & 9s   & \textbf{0.01s} \\
\bottomrule
\end{tabular}
}%
\vspace{-2mm}
\end{table}

\paragraph{Baselines.} We benchmark our proposed system against several established models in both 3D and 2D image generation and editing: \emph{SDFusion} \cite{cheng2023sdfusion}, a latent diffusion-based model for generating 3D shapes through signed distance functions; \emph{AutoSDF}\cite{autosdf2022}, which leverages a 3D autoencoder and BERT \cite{BERT} to generate signed distance fields; \emph{shap-E}\cite{Shape2023}, designed for generating conditional 3D implicit functions; LLaMA-Mesh \cite{llamamesh}, enables 3D generation from LLM by tokenizing the vertices and faces of the 3D meshes; \emph{LGM}\cite{tang2024lgm} (Large Multi-View Gaussian Model), which combines a pretrained 2D diffusion model with a large Gaussian model to generate 3D Gaussian representation of the model; \emph{Stable Diffusion}\cite{stablediffusion}, a leading image generation model, used as a baseline for evaluating image synthesis from text input; \emph{TurboEdit}\cite{turboedit}, a diffusion-based model that edits 2D  based on text prompts, serving as a reference for comparing the 2D editing capabilities of our system against conventional 2D approaches; and \emph{InstructPix2Pix}\cite{instructpix2pix}, a text-driven 2D image editing model with stable diffusion backbone.

\subsection{Experimental Details}
Proc3D enables any LLM to generate 3D objects using our novel PCG representation. To test its flexibility, we evaluate four models: GPT-4o and LLaMA 3-70B using in-context learning (ICL), and fine-tuned variants of LLaMA 3 with 8B and 70B parameters. For ICL, 20 examples were retrieved via BM25 ranking \cite{Amati2009BM25} from the instruction-graph examples, and were passed into the model, together with instruction prompts. For a robust evaluation, we used QLORA to fine-tune LLaMA-3 8B and LLaMA-3 70B for 5 epochs using our curated instruction-graph dataset, with a 90\%/10\% split for training and validation. 

To comprehensively assess performance, we sampled 100 in-distribution (IID) and 100 out-of-distribution (OOD) unique prompts from GPT-4o, covering object categories both seen and unseen during training. These prompts were used consistently across all quantitative evaluations. This approach allowed us to fairly compare the models' performance and evaluate their ability to generalize to new prompts.
\begin{figure}
    \centering
\includegraphics[width=0.98\linewidth]{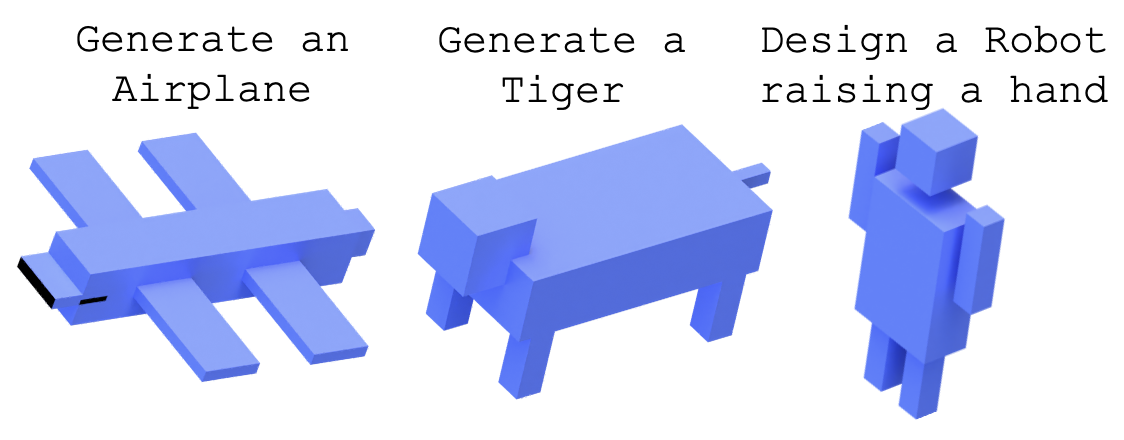}
    \caption{\textbf{Generation out of training classes.} 
With training data limited to chairs, tables, beds, and storage, Proc3D can generate complex 3D categories that are not in the training examples. }
    \label{fig:out-of-training}
\end{figure}
\subsection{Quantitative Results}
\paragraph{3D Representation.}
The PCG representation is developed to address the inefficiencies of previous LLM-compatible representation such as the blender's geometric nodes, Infinigen \cite{infinigen2023infinite} abstraction, and the python code in Blender\cite{BlenderGPT, scenecraft, L3goYamada2024}. In Table \ref{tab:representation_comparison}, we compare generation of these representations with 100 example prompts using GPT-4o with 128K context length. Results show that, by optimizing the representation of the 3D model, compile rates significantly increases, enhancing the effectiveness and efficiency of procedural generation. This advancement not only ensures more reliable outputs but also supports the use of a higher number of ICL examples for non-finetuned LLMs. We also compare these representations using more recent reasoning-oriented LLM, as shown in Table~\ref{tab:compile_efficiency_reasoning} of the supplementary material, to highlight their impact on compilation performance and generalization.

\paragraph{3D Generation.}
We present a quantitative comparison of Proc3D with the baselines in Table \ref{tab:ULIP_comparison}. Proc3D achieves the highest ULIP score, demonstrating superior alignment between text prompts and the generated 3D models compared to baseline methods. This result suggests that Proc3D excels at interpreting and translating text descriptions into accurate 3D representations, capturing detailed geometry and structural details with higher precision than the baselines. Notably, Proc3D with GPT-4o achieves generation times similar to those of the highest-quality previous work \cite{cheng2023sdfusion} while enabling real-time editing of the 3D model, which is 400x more efficient than the baselines which require regenerating the 3D model and optimizing input prompts to produce the desired 3D output.
\begin{table}[b!]
\vspace{-4mm}
\centering
\caption{\textbf{Evaluation of few-shot and fine-tuned models on both in-training (In-tr.) and out-of-training (Out-tr.) data categories.} Fine-tuned models achieve higher compile rates, but tend to lack the generalization capabilities of non-fine-tuned models. However, they consistently follow the text prompts and generate higher quality models, resulting in higher ULIP scores compared to few-shot models.}
\vspace{1mm}
\scriptsize  
\begin{tabular}{lllllll}
\toprule
\textbf{Model} & \multicolumn{2}{c}{\textbf{Compile Rate↑}}  & \multicolumn{2}{c}{\textbf{Similarity Measure↑}} & \multicolumn{2}{c}{\textbf{ULIP↑}} \\ 
\cmidrule(lr){2-3} \cmidrule(lr){4-5} \cmidrule(lr){6-7}
& \textbf{In-tr.} & \textbf{Out-tr.} & \textbf{In-tr.} & \textbf{Out-tr.} & \textbf{In-tr.} & \textbf{Out-tr.} \\ 
\midrule
\textbf{Few-Shot ICL} & & & & & & \\
GPT4o          & 0.89 & 0.86 & \textbf{0.58} & 0.97 & 0.14 & 0.10 \\
llama3-70B     & 0.75 & 0.68 & 0.56 & 0.86 & 0.13 & 0.07 \\
o3         & \textbf{ 0.94} & \textbf{0.92 }& 0.62 & \textbf{0.98} & \textbf{0.15} & \textbf{0.13} \\
DeepSeek-V3  & 0.92 & 0.91 & \textbf{0.64} & \textbf{0.98} & 0.14 & 0.11 \\
\midrule
\textbf{Finetuned} & & & & & & \\
llama3-8B      & \textbf{0.98} & \textbf{0.97} & 0.46 & 0.49 & 0.13 & 0.06 \\
llama3-70B     & \textbf{0.98} & 0.96 & 0.50 & 0.58 & \textbf{0.15} & 0.08 \\
\bottomrule
\end{tabular}
\normalsize
\vspace{-6mm}
\label{tab:ulip_and_compile}
\end{table}

\paragraph{\textbf{Ablation Studies.}} 

In Table~\ref{tab:ulip_and_compile}, we compare LLMs for 3D model generation using Compile Rate and Similarity Measure across both few-shot ICL and fine-tuned settings. Fine-tuned models achieve higher Compile Rates, reflecting better syntactic understanding of PCG, but tend to overfit, leading to lower Similarity on out-of-training categories. Despite this, their higher ULIP scores indicate more semantically accurate outputs. For instance, LLaMA-8B underperforms GPT-4o in ULIP, highlighting the impact of overfitting on generalization. Further ablation to assess the effect of ICL example count on PCG quality is shown in the supplementary material.
\begin{figure}
    \centering
    \includegraphics[width=0.98\linewidth]{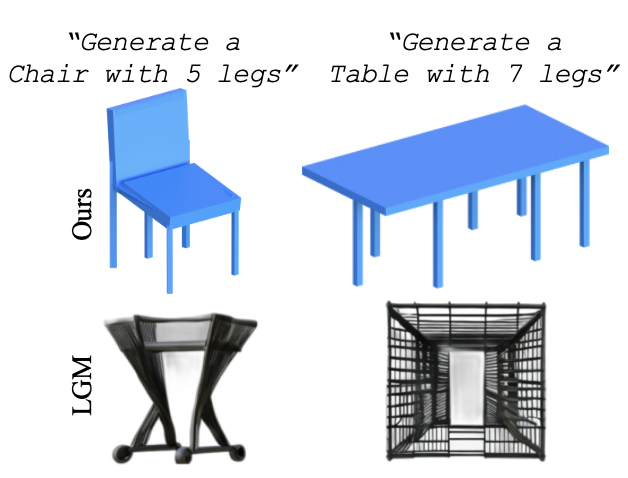}
    \caption{\textbf{Unconventional Object Generation}. Proc3D can follow the text prompt to generate unconventional objects better than LGM\cite{tang2024lgm}}
    \label{fig:unconventional}
    \vspace{-5mm}
\end{figure}
\paragraph{3D Editing.}
Quantitative comparisons of 3D editing methods are challenging due to the lack of publicly available benchmarks and accessible implementations of alternative methods. For example, we could not access the code for methods like ParSEL \cite{parsel2024}, which reportedly take 30 to 500 seconds per mesh edit. In contrast, as shown in Table \ref{tab:ULIP_comparison}, and in the Figure \ref{fig:compare_edits} 
Proc3D achieves edits in under 10 milliseconds, significantly outperforming these methods (\textbf{over 400x speedup}). This performance enables real-time 3D object modifications, making Proc3D a highly practical tool for dynamic 3D editing workflows.

\subsection{Qualitative Results}
\paragraph{3D Generation.} 
We present qualitative results of 3D generation in Figure \ref{fig:results}, which demonstrate Proc3D's ability to generate 3D models from text inputs. Figure \ref{fig:qualitative_compare} further shows the flexibility of Proc3D in generating a wide varieties of 3D shapes, from everyday objects like chairs and tables to abstract forms like the Olympic rings. Compared to baseline models, Proc3D exhibits a clear advantage in maintaining geometric consistency and capturing precise details. 

We also show out-of-distribution generation examples in Figure \ref{fig:out-of-training}. Due to the nature of our training dataset, which consists solely of cube primitives, the generated objects, including more complex ones like Tigers or Airplanes, are still composed of cube-based structures. In particular, generating such out-of-distribution objects is not directly possible with traditional code-representation methods \cite{3dpremise, L3goYamada2024} or vertices-faces generation from LLM \cite{llamamesh} without our simplified PCG language, which enables LLM to effectively reason about these models. To further illustrate the advantages of our approach, we show the generation of unconventional objects in Figure \ref{fig:unconventional} and compare them against state-of-the-art diffusion and Gaussian representations \cite{tang2024lgm}. Our method consistently outperforms these baselines, demonstrating superior capability in generating even abstract shapes by following the user's prompts.

\paragraph{3D Editing.} 

We present edit results of our system in Figure \ref{fig:results}, and several examples in the appendix, where Proc3D edits the generated 3D models either manually with sliders or checkboxes, or with  natural language prompts. The right side of each example focuses on editing capabilities, allowing users to make precise modifications, such as removing components or adjusting dimensions, while preserving the structural integrity of the models. Compared to the other baselines in Figure \ref{fig:qualitative_compare}, which use representations that are static and not inherently editable, Proc3D allows direct editing of the generated 3D models. This ability to modify the generated shapes directly, as shown in the figure, provides users with greater flexibility and control over the design process.

\begin{figure}
\vspace{-4mm}
    \centering
    \includegraphics[width=0.95\linewidth]{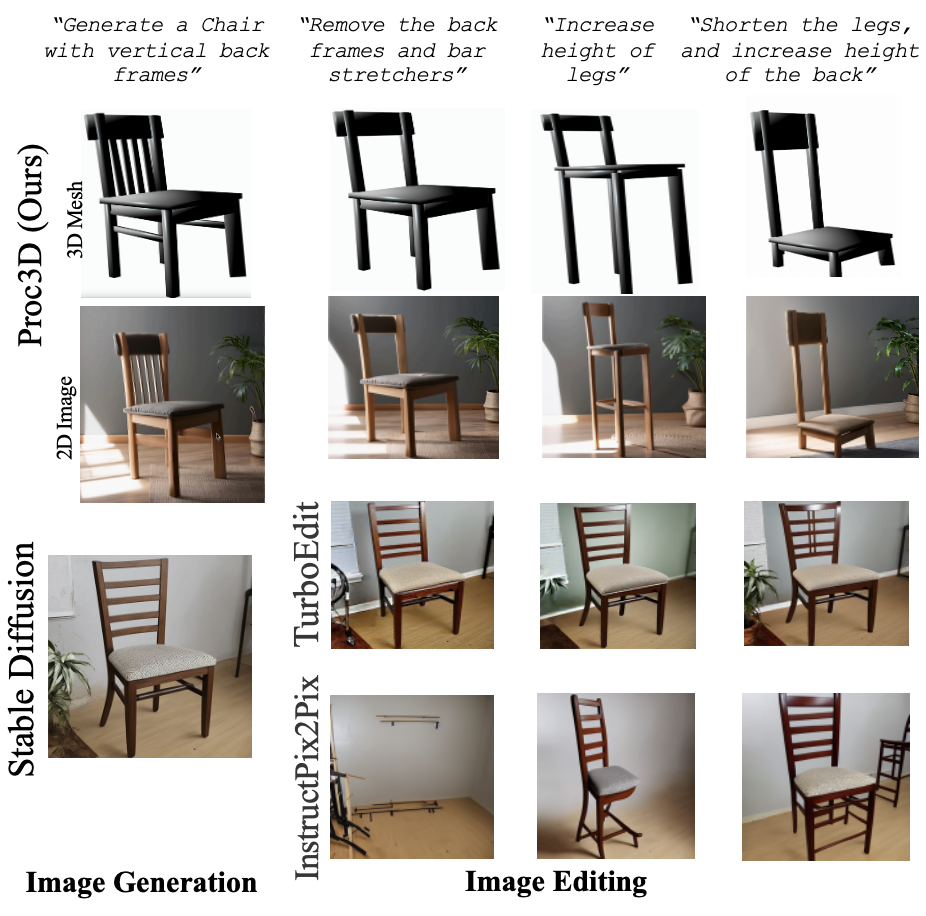}
    \caption{\textbf{Comparison of Image Generation and Editing Capabilities}. The top two rows is our system, Proc3D, which generates 3D meshes and corresponding 2D  from text prompts, followed by interactive editing capabilities on the generated 3D, and consequently the 2D image. The bottom rows compare various baseline methods: Stable Diffusion \cite{stablediffusion} for image generation, TurboEdit\cite{turboedit} and Instruct-pix2pix\cite{instructpix2pix} for image editing.
    }
    \vspace{-2mm}
    \label{fig:2D_generation_edit}
\end{figure}

Editing 3D models can also significantly benefits 2D image generation and editing, as projections of 3D models can guide the image generation process \cite{zhang2023adding}. Proc3D offers fine-grained 3D model editing capabilities, leveraging enhanced 3D controllability to generate  more closely aligned with user requests or views. We compare our method with traditional image generation and editing methods in Figure \ref{fig:2D_generation_edit}, showing that our system produces  that better meet users' specifications.

\section{Conclusion and Discussion}
In this work, we introduced Proc3D, a framework that empowers LLMs to generate and edit 3D objects through our novel Procedural Compact Graph (PCG) representation. PCG provides an efficient representation of 3D models, reducing complexity by over 4-10 times compared to traditional methods. Qualitative and Quantitative results shows that Proc3D outperforms state-of-the-art models, such as the Large Multi-view Gaussian Model (LGM) \cite{tang2024lgm}, in accurately following text instructions for unconventional objects (Figure \ref{fig:unconventional}) and atypical designs, such as the Olympic rings (Figure \ref{fig:qualitative_compare}). This demonstrates Proc3D’s superior ability to interpret and respond to diverse text prompts. Additionally, our method achieves the highest ULIP scores, indicating strong alignment between input prompts and generated 3D models, surpassing baselines like SDFusion, AutoSDF, LLaMA-Mesh and shap-E. 

\paragraph{\textbf{Limitation and Future Direction.}} Proc3D demonstrates strong capabilities in both the generation and editing of 3D models and scenes using a procedural graph representation. The model generalizes well to unseen structures and shapes, including non-trivial forms such as the Olympic rings composed of toroidal elements (Figure~\ref{fig:out-of-training}). This suggests that the learned representation is not tightly coupled to specific geometric templates. However, the diversity of current training data remains limited, as existing datasets predominantly feature low-complexity primitives. In the future, we aim to further enrich the diversity and complexity of procedural training data to support broader generalization across categories, primitives, and scene compositions.

\bibliography{ref}
\newpage
\appendix
\onecolumn
\section{Procedural Compact Graph (PCG)}

In section 3 of the main paper, we presented, and show a full example of the PCG (see Figure \ref{fig:pcg_example}), a representation that abstracts the complexity of software-specific representations of procedural 3D modeling. We now show the corresponding Python code using Infinigen \cite{infinigen2023infinite} transpiler in Figure \ref{fig:table_code}. The Blender's native code is too long to show here. As shown, the Infinigen abstraction of the native blender's code is significantly longer in token size compared to the concise format of PCG, highlighting the efficiency and simplicity that PCG offers. By reducing code length and complexity, PCG not only makes the modeling process more manageable but also enhances the capability of Large Language Models (LLMs) to generate and interpret 3D models effectively.

\paragraph{PCG Example.} The PCG in Figure \ref{fig:pcg_example} of the main paper which can be interpreted and rendered by Proc3D, defines the creation of a table model. The graph begins by specifying input parameters such as \textit{table\_width}, \textit{table\_length}, \textit{leg\_height}, and \textit{leg\_radius}. It then defines the legs using \textit{cylinder} nodes, positioned at the four corners using a \textit{rectangle} node and a \textit{point instance} node, with a \textit{transform} node adjusting their placement. The tabletop is created using a \textit{rectangle} node, beveled with a \textit{fillet} node, smoothed using a \textit{fill} node, and given thickness via an \textit{extrude} node. Finally, the legs and tabletop are combined using a \textit{join} node to output the completed model.


\paragraph{Extracting the PCG from 3D Meshes}
Extraction of PCG from 3D meshes involves identifying and encoding the transformations, geometric properties, and spatial relationships inherent to the model. This process requires careful analysis of each mesh element to abstract its procedural structure, allowing the PCG to represent the model in a simplified yet comprehensive format. The tranformation is highlighted below:
\paragraph{Transformation.}

Given a 3D mesh \( \mathcal{M} \) of a leaf node in the hierarchical mesh dataset \cite{mo2019partnet} represented by vertices \( \mathbf{V} = \{ \mathbf{v}_1, \mathbf{v}_2, \dots, \mathbf{v}_N \} \), we extract the translation, rotation, and scaling transformations.
\subparagraph{Translation.}
The translation vector \( \mathbf{T} \) is computed as the centroid of the mesh:

\[
\mathbf{T} = \frac{1}{N} \sum_{i=1}^{N} \mathbf{v}_i
\]

\subparagraph{Rotation.}

The rotation is extracted by computing the principal axes through PCA. First, we compute the covariance matrix:
\[
C = \frac{1}{N} \sum_{i=1}^{N} \mathbf{v}_i' \mathbf{v}_i'^T
\]
where $\mathbf{v}_i' = \mathbf{v}_i - \mathbf{c}$, and the eigenvectors \( \hat{\mathbf{d}}_1, \hat{\mathbf{d}}_2, \hat{\mathbf{d}}_3 \) of \( C \) form the rotation matrix \( R \) = $\begin{bmatrix} \hat{\mathbf{d}}_1 & \hat{\mathbf{d}}_2 & \hat{\mathbf{d}}_3 \end{bmatrix}$

Then, we convert the rotation matrix \( R \) into Euler angles \( (\phi, \theta, \psi) \), which represent rotations around the x-, y-, and z-axes, respectively. 

\subparagraph{Scaling.}
The scaling factors \( l_1, l_2, l_3 \) are the extents of the bounding box along the principal axes (given by the eigenvectors). These represent the lengths along each principal axis.
\[
l_1 = \text{extent along } \hat{\mathbf{d}}_1, \quad \\ l_2 = \text{extent along } \hat{\mathbf{d}}_2, \quad \\ l_3 = \text{extent along } \hat{\mathbf{d}}_3
\]

\section{Node Definitions}
\label{sec:node_definitions}
Following \cite{infinigen2023infinite}, we identified and translated 79 unique nodes from Blender's geometry nodes into intuitive text representations that are both human-readable and machine-interpretable. Some key node types include:

\begin{itemize}
    \item \textbf{Input}: Defines input parameters for PCG. These parameters are semantically defined by the variables corresponding to the specific parts they modify.
    \item \textbf{Mesh Primitives}: Basic geometric shapes that serve as building blocks for models.
    \begin{itemize}
        \item \textit{cube}: Creates a unit cube centered at the origin, with default sides of length 1 along each axis.
        
        \item \textit{cylinder}: Generates a cylinder aligned along the z-axis with specified radius and depth
       
        \item \textit{sphere}: Produces a sphere centered at the origin with a defual radius value of 1
    \end{itemize}
    \item \textbf{Transform}: Operations to manipulate geometry.
    \begin{itemize}
        \item \textit{T(x, y, z)}: Moves geometry by the specified vector $\mathbf{t} = (x, y, z)$. 
        \item \textit{R(rx, ry, rz)}: Rotates geometry using Euler angles around the x-, y-, and z-axes by angles $r_x$, $r_y$, and $r_z$, respectively. 
        \item \textit{S($s_x$, $s_y$, $s_z$)}: Scales geometry along the specified axes. 
    \end{itemize}
    \item \textbf{Geometry Operations}: Combine or modify meshes based on spatial relationships.
    \begin{itemize}
        \item \textit{join}: Combines the output of two or more geometry nodes into a single unified structure, represented as the union \(\mathcal{M} = \mathcal{M}_1 \cup \mathcal{M}_2 \cup \dots \)
        \item \textit{switch}:  Allows conditional selection between different geometries or operations based on a boolean input. This enables toggling parts of the model on or off.
        \item \textit{combine}: Merges separate scalar values into a vector. For instance, combining individual $x$, $y$, and $z$ components into a single position vector.
        \item \textit{extrude}: Extends geometry along its normals by a specified distance.
    \end{itemize}
    \item \textbf{Math Operations}: Perform arithmetic operations on numerical inputs or attributes, often used for parametric controls.
    \begin{itemize}
        \item \textit{add}: Computes the sum of two inputs, such as adding two scalar values, vectors, or attributes.

        \item \textit{subtract}: Subtracts one input from another, typically used for adjusting positions, sizes, or other attributes.
        
        \item \textit{multiply}: Multiplies two inputs, often used for scaling or other proportional adjustments.

        \item \textit{divide}: Divides one input by another, useful for normalizing values or adjusting parameters.
    \end{itemize}
    \item \textbf{Output}: Defines the output of the procedural graph.
\end{itemize}
\section{Experiments}

\begin{figure*}
    \centering
    \includegraphics[width=1\linewidth]{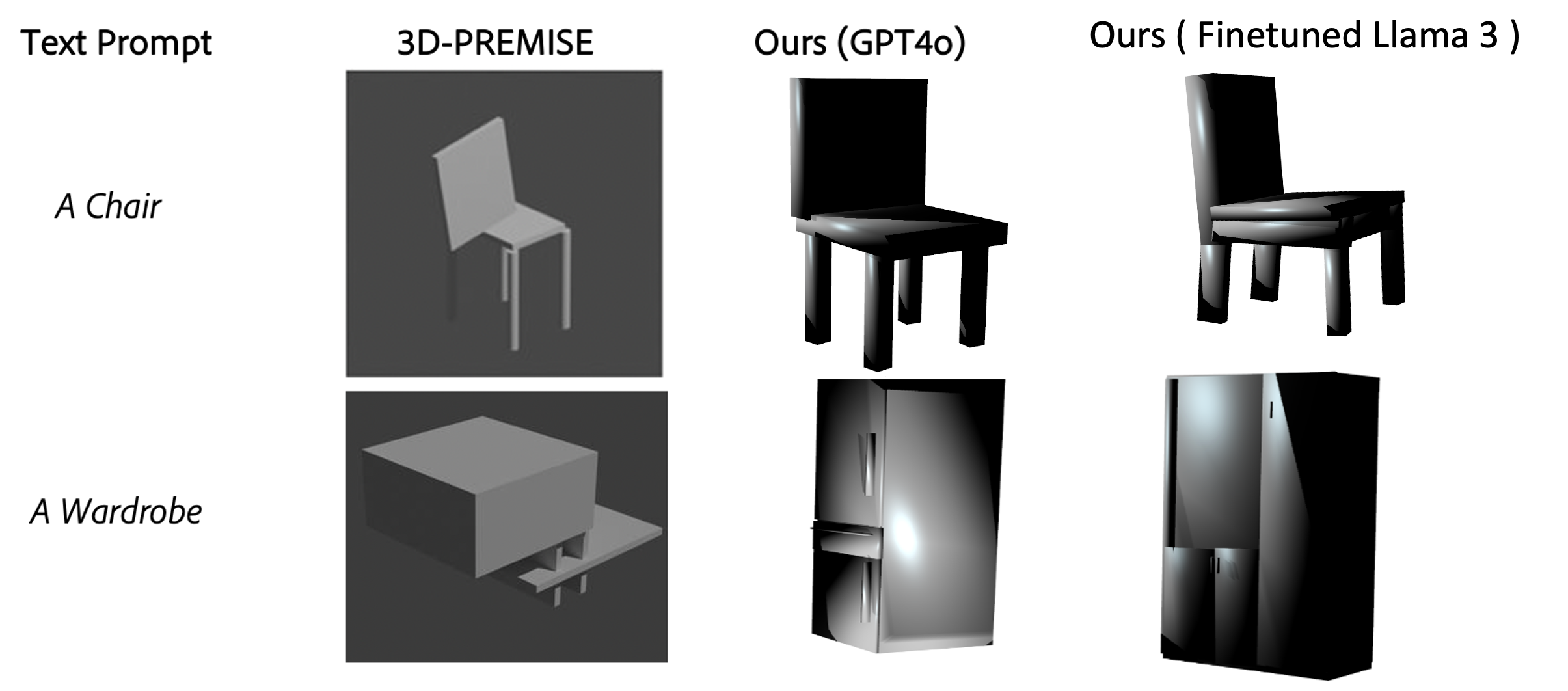}
    \caption{\textbf{Qualitative Comparison with LLM-Driven 3D method}. Our method shows improvement over 3DPremise \cite{3dpremise}, which uses native blender's code for 3D object generation.}
    \label{fig:compare_3dpremise}
\end{figure*}

\begin{figure*}[b!]
    \centering
    \includegraphics[width=1\linewidth]{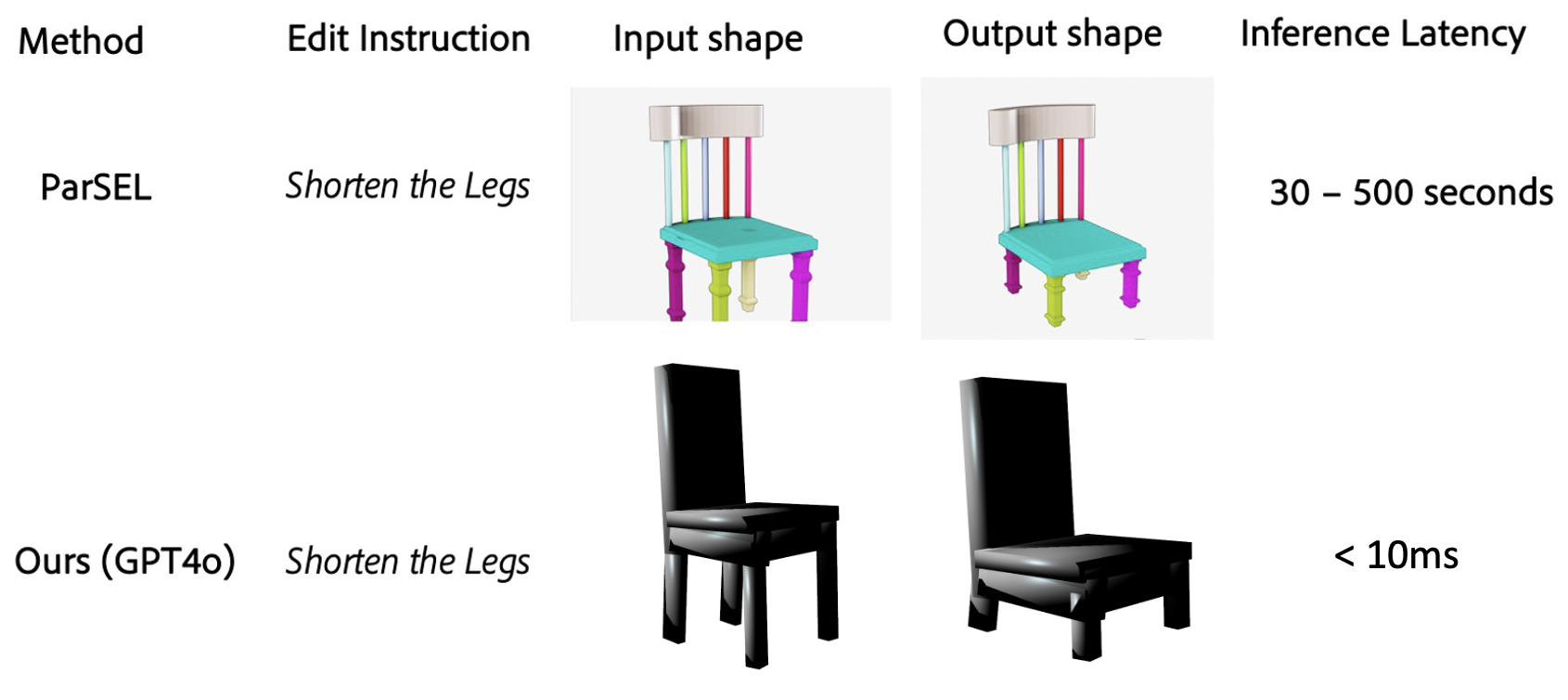}
    \caption{\textbf{Quatitative Comparison with LLM-Driven 3D Editing method}. Our method shows improvement over ParSEL \cite{parsel2024} which directly edits 3D mesh representation.}
    \label{fig:compare_edits}
\end{figure*}

\textbf{3D Generation}. We show qualitative comparison with 3DPremise \cite{3dpremise}, an LLM-driven 3D generation method in Figure \ref{fig:compare_3dpremise}, and PerSel for 3D editing in Figure \ref{fig:compare_edits}.  Our method demonstrate improvements both in the quality of the generated 3D models and the time required to edit them. Proc3D enables the effectiveness in producing more detailed and realistic 3D models while enabling faster and more intuitive editing workflows compared to existing methods.

\begin{figure*}[t]
\centering
\includegraphics[width=1\textwidth]{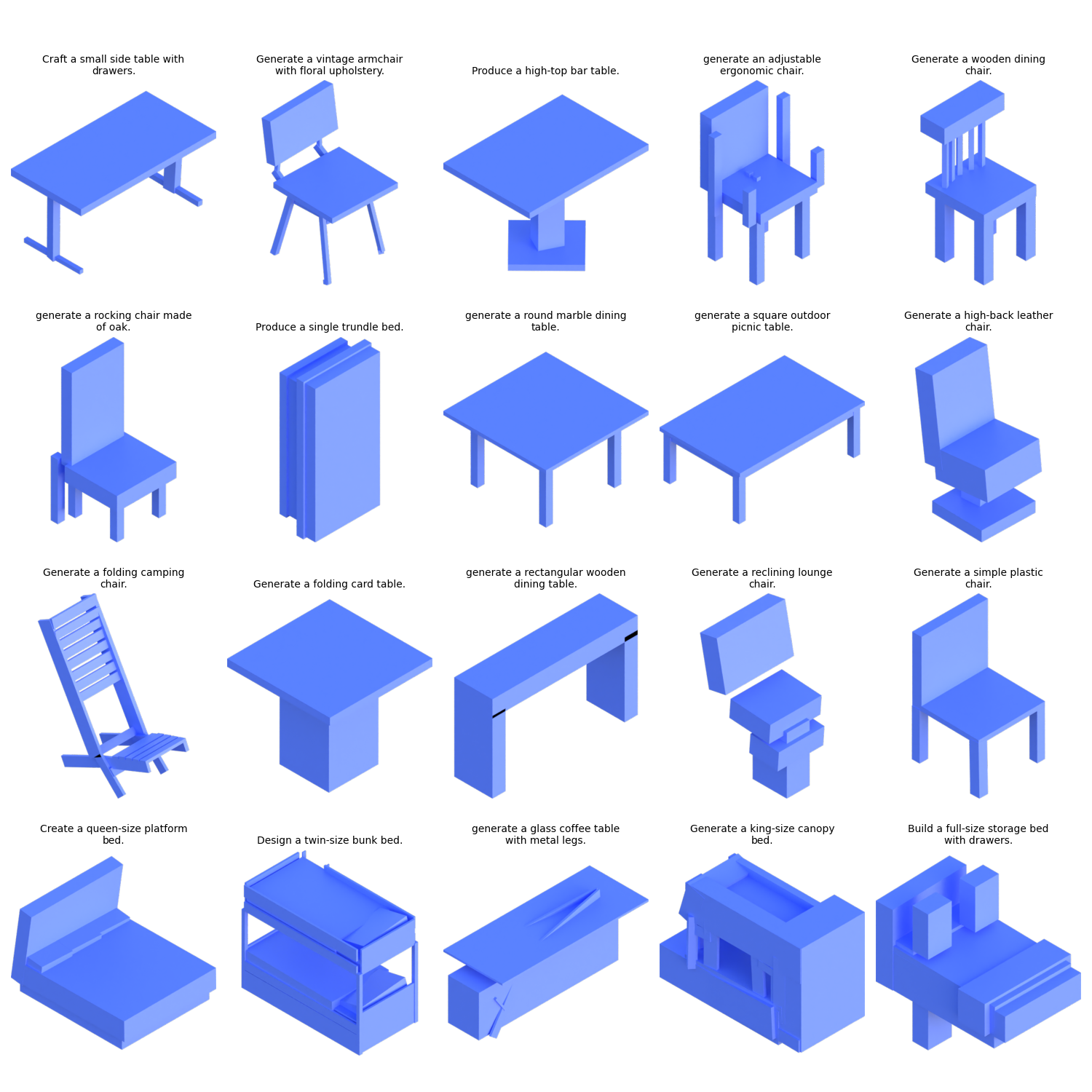} 
\caption{\textbf{Qualitative Examples of the Proc3D system from uniquely generated text prompts}. These are examples of the randomly sampled prompts used for all quantitative experiments in the main paper.}
\label{fig:samples}
\vspace{-3mm}
\end{figure*}

\begin{figure*}[t]
\centering
\includegraphics[width=1\textwidth]{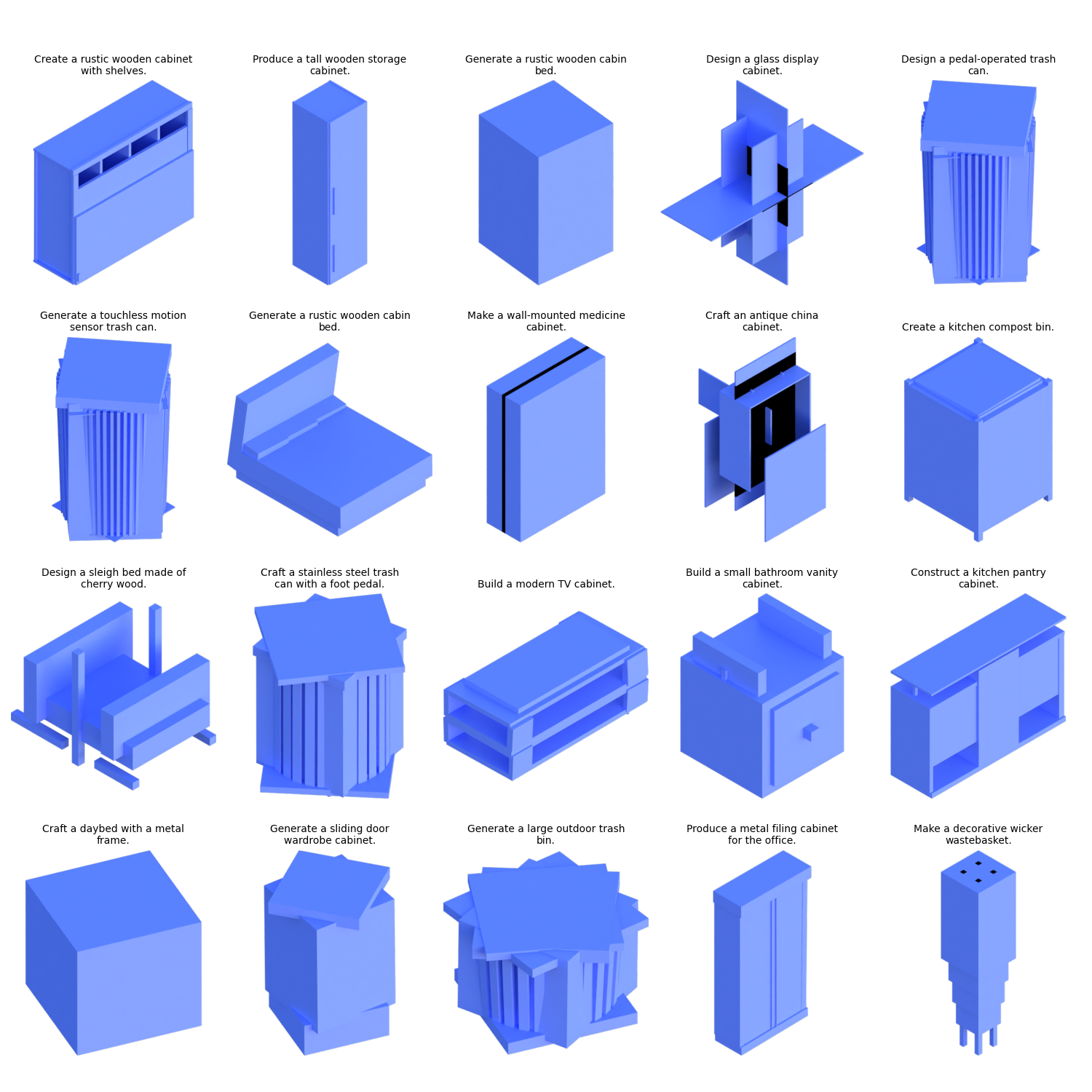} 
\caption{\textbf{Qualitative Examples of the Proc3D system from uniquely generated text prompts}. These are examples of the randomly sampled prompts used for all quantitative experiments in the main paper.}
\label{fig:samples 2}
\vspace{-3mm}
\end{figure*}

We further show several examples of the 3D models generated by Proc3D in Figure \ref{fig:samples}, and Figure \ref{fig:samples 2}

\begin{figure*}
    \centering
    \includegraphics[width=0.85\linewidth]{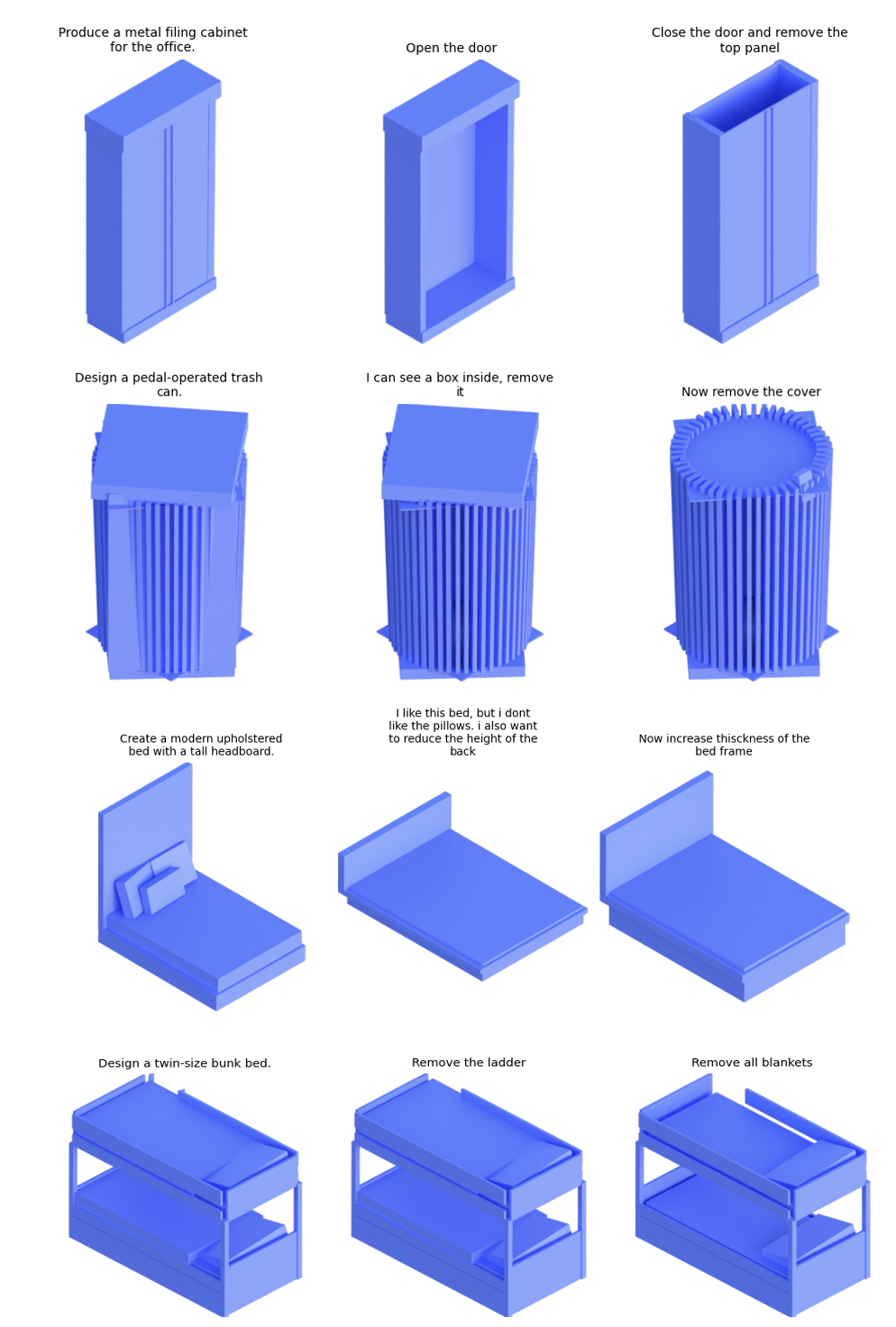}
    \caption{\textbf{Qualitative Editing Results of Proc3D}. Proc3D can precisely edit the generated 3D models}
    \label{fig:editing_1}
\end{figure*}

\begin{figure*}
    \centering
    \includegraphics[width=0.85\linewidth]{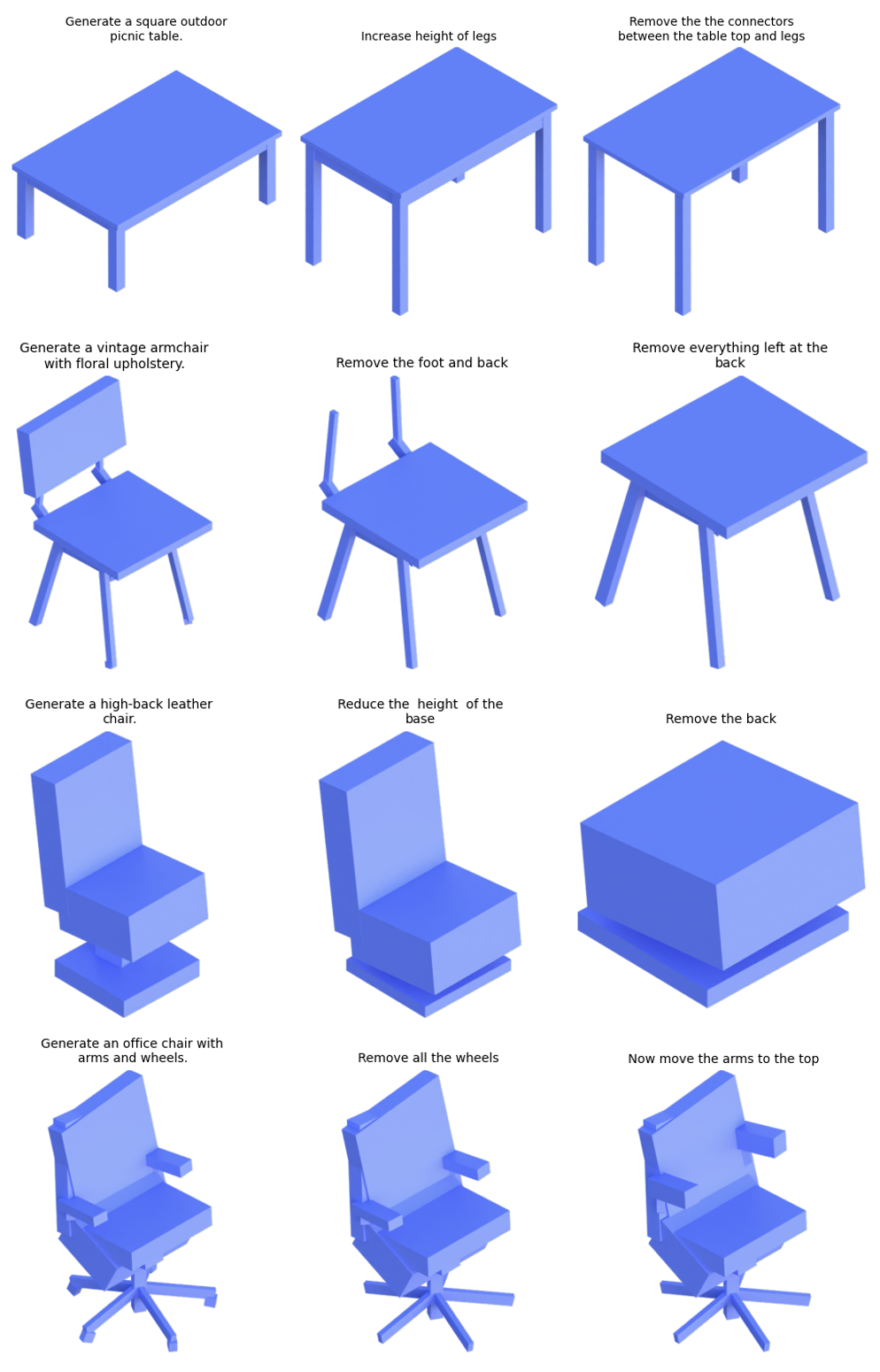}
    \caption{\textbf{Qualitative Editing Results of Proc3D}. Proc3D can precisely edit the generated 3D models}
    \label{fig:editing_2}
\end{figure*}

\begin{figure*}
    \centering
    \includegraphics[width=0.85\linewidth]{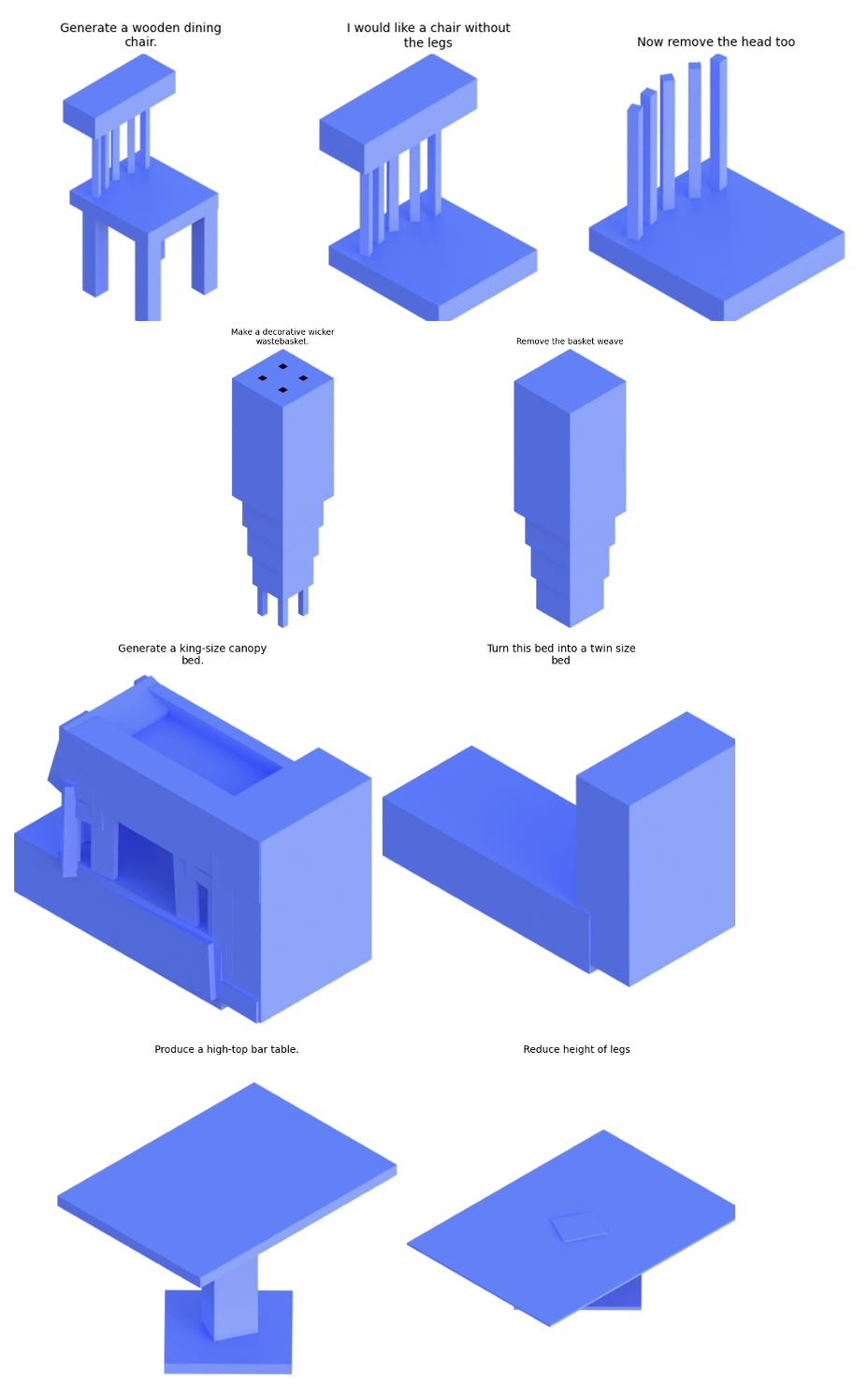}
    \caption{\textbf{Qualitative Editing Results of Proc3D}. Proc3D can precisely edit the generated 3D models}
    \label{fig:editing_3}
\end{figure*}

\textbf{3D Editing}. Figures \ref{fig:editing_1}, \ref{fig:editing_2}, and \ref{fig:editing_3} demonstrate the process of editing the generated 3D models, showing how different modifications can be applied to refine the 3D structure, adjust components, and enhance their overall design. These figures highlight the effectiveness of our approach in enabling intuitive and precise 3D editing. 

\begin{figure*}[b!]
    \centering
    \includegraphics[width=0.85\linewidth]{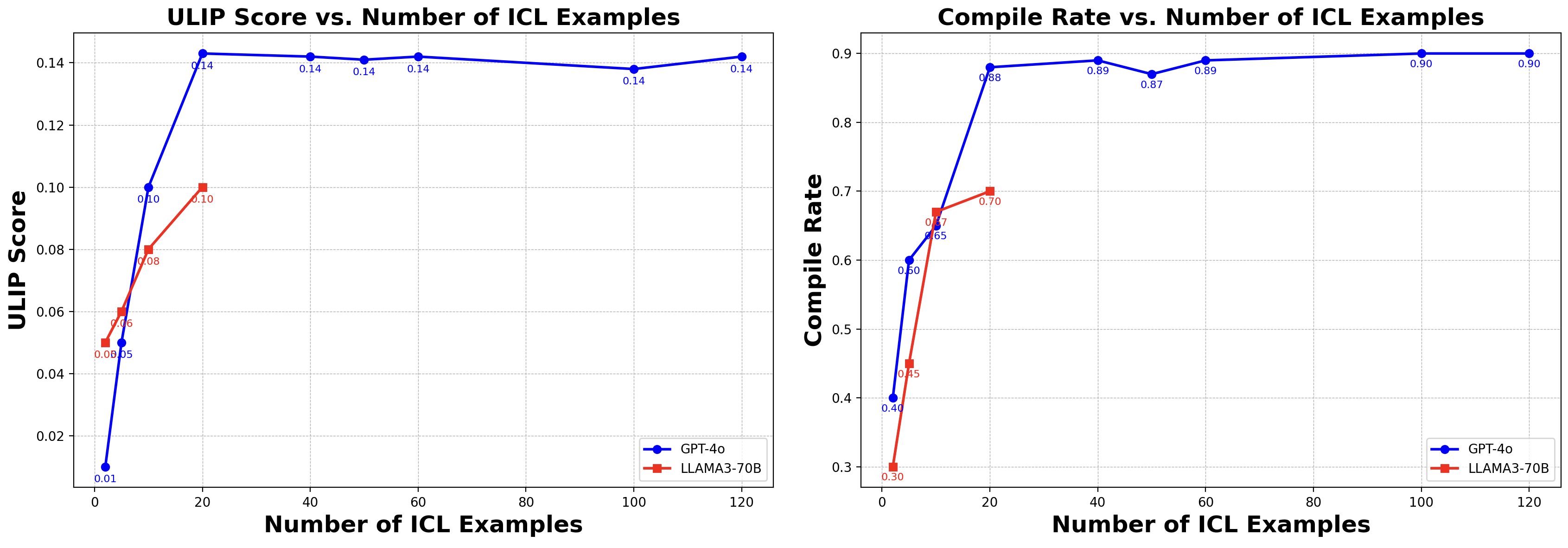}
    \caption{\textbf{Ablation Study on ICL Examples for 3D Model Generation}. Results show performance improvements up to 20 examples, with no major improvement thereafter, indicating an optimal threshold for model performance in PCG generation. }
    \label{fig:ablation-ICL}
\end{figure*}

\begin{table}[t]
\centering
\caption{\textbf{Compilation Efficiency Comparison Across Reasoning Models.}
We compare different procedural representations evaluated using reasoning-oriented LLMs ( o3, and DeepSeek-V3). 
Our PCG representation achieves the highest compile rate (CR) while substantially reducing token count and generation time on both o3 and DeepSeek-V3 reasoning models.}
\vspace{-2mm}
\resizebox{\linewidth}{!}{
\begin{tabular}{lcccccc}
\toprule
\multirow{3}{*}{\textbf{Representation}} & 
\multicolumn{3}{c}{\textbf{o3 (Reasoning Model)}} & 
\multicolumn{3}{c}{\textbf{DeepSeek-V3 (Reasoning Model)}} \\
\cmidrule(lr){2-4} \cmidrule(lr){5-7}
 & \textbf{CR (\%)} & \textbf{Tokens} & \textbf{Time (s)} 
 & \textbf{CR (\%)} & \textbf{Tokens} & \textbf{Time (s)} \\
\midrule
Blender Geometry Nodes & 0  & 6035 & 420 & 0  & 6027 & 279 \\
Infinigen DSL           & 15 & 3270 & 238 & 9  & 3190 & 225 \\
Blender Python          & 65 & 2660 & 151 & 63 & 2585 & 194 \\
\textbf{PCG (Ours)}     & \textbf{94} & \textbf{655} & \textbf{73} & \textbf{92} & \textbf{689} & \textbf{61} \\
\bottomrule
\end{tabular}}
\vspace{-3mm}
\label{tab:compile_efficiency_reasoning}
\end{table}

\paragraph{\textbf{Ablation Studies.}} 

We further ablate LLaMA-3 70B and GPT-4o to assess the effect of ICL example count on PCG quality. Due to its 8K token limit, LLaMA-3 was constrained to 20 ICL examples from chair and table categories (avg. 302 tokens each), while GPT-4o’s 128K context allowed up to 300. Results in Figure~\ref{fig:ablation-ICL} show rapid performance gains up to 20 examples, after which improvements plateau, suggesting an effective ICL threshold. These findings reinforce that the PCG representation allows strong compositional reasoning from limited demonstrations.

\section{Conclusion}
We presented Proc3D, a method for generation and real-time editing of 3D models. Our approach advances generation of 3D models from LLMs by generating higher-quality models and enabling faster, more intuitive editing workflows. Despite limitations due to dataset constraints and the use of a single cube primitive, Proc3D represents a step toward more efficient and user-friendly 3D modeling. Future work will aim to address these challenges to enhance the system's robustness and applicability.


\begin{figure*}[ht]
\centering
\begin{minipage}{\textwidth}
\begin{lstlisting}[style=mypython, label={lst:partnet_graph}]
def extract(obj_filename):
    obj = load_object(obj_filename)
    boxes = obj.boxes(leafs_only=True)

    current_tokens = {"inputs": {"Geometry": "cube"}}
    created_vectors = {"vector_counts": {}}
    done = {}
    switches = {}

    for node in obj.depth_first_traversal():
        if node.is_leaf:
            box = node.box
            box[:3] = coord_rot @ box[:3]
            box[6:9] = coord_rot @ box[6:9]
            box[9:] = coord_rot @ box[9:]
            
            T, R, S = extract_transformations(box)
            names = clean_path(node.full_label)

            # Traverse and build hierarchy graph
            current = done
            for i, name in enumerate(names):
                if name not in current:
                    current[name] = [] if i == len(names)-2 else {}
                current = current[name]

            token, graph_str = attach_part(name, (T, R, S), current_tokens, created_vectors, switches)
            current.append(token)

    switch_str, changed_tokens = join_all_switches(switches, current_tokens)
    token, graph_str2 = recursively_join(done, changed_tokens, current_tokens)

    token, graph_str3 = expose_rotation(global_name, token, current_tokens)

    output =format_output(token, switch_str + graph_str2, current_tokens, created_vectors)
    save_graph(output, obj_filename)
\end{lstlisting}
\end{minipage}
\caption{{Extracting Transformation and Composing Graph from PartNet Dataset}}
\label{fig:psuedo_code}
\end{figure*}
\newpage

\begin{figure*}[ht]
\centering
\begin{minipage}{\textwidth}
\begin{lstlisting}
# Example Python code
import bpy
import mathutils
from numpy.random import uniform, normal, randint
from infinigen.core.nodes.node_wrangler import Nodes, NodeWrangler
from infinigen.core.nodes import node_utils
from infinigen.core.util.color import color_category
from infinigen.core import surface

def geometry_nodes(nw: NodeWrangler):
    # Code generated using version 2.6.5 of the node_transpiler

    group_input = nw.new_node(Nodes.GroupInput,
        expose_input=[('NodeSocketGeometry', 'Geometry', None),
            ('NodeSocketFloatDistance', 'TableWidth', 2.0000),
            ('NodeSocketFloatDistance', 'TableHeight', 2.0000),
            ('NodeSocketFloatDistance', 'TopCurve', 0.2500),
            ('NodeSocketFloat', 'TopTickness', 1.0000),
            ('NodeSocketFloatDistance', 'LegRadius', 1.0000),
            ('NodeSocketFloatDistance', 'LegHeight', 2.0000)])
    
    subtract = nw.new_node(Nodes.Math, input_kwargs={0: group_input.outputs["TableWidth"]}, attrs={'operation': 'SUBTRACT'})
    
    subtract_1 = nw.new_node(Nodes.Math, input_kwargs={0: group_input.outputs["TableHeight"]}, attrs={'operation': 'SUBTRACT'})
    
    quadrilateral_1 = nw.new_node(Nodes.Quadrilateral, input_kwargs={'Width': subtract, 'Height': subtract_1})
    
    cylinder = nw.new_node(Nodes.Cylinder,
        input_kwargs={'Radius': group_input.outputs["LegRadius"], 'Depth': group_input.outputs["LegHeight"]})
    
    instance_on_points = nw.new_node(Nodes.InstanceOnPoints, input_kwargs={'Points': quadrilateral_1, 'Instance': cylinder.outputs["Mesh"]})
    
    divide = nw.new_node(Nodes.Math,
        input_kwargs={0: group_input.outputs["LegHeight"], 1: -2.0000},
        attrs={'operation': 'DIVIDE'})
    
    combine_xyz = nw.new_node(Nodes.CombineXYZ, input_kwargs={'Z': divide})
    
    transform_geometry = nw.new_node(Nodes.Transform, input_kwargs={'Geometry': instance_on_points, 'Translation': combine_xyz})
    
    quadrilateral = nw.new_node(Nodes.Quadrilateral,
        input_kwargs={'Width': group_input.outputs["TableWidth"], 'Height': group_input.outputs["TableHeight"]})
    
    fillet_curve = nw.new_node(Nodes.FilletCurve,
        input_kwargs={'Curve': quadrilateral, 'Count': 20, 'Radius': group_input.outputs["TopCurve"]})
    
    fill_curve = nw.new_node(Nodes.FillCurve, input_kwargs={'Curve': fillet_curve}, attrs={'mode': 'NGONS'})
    
    extrude_mesh = nw.new_node(Nodes.ExtrudeMesh,
        input_kwargs={'Mesh': fill_curve, 'Offset Scale': group_input.outputs["TopTickness"]})
    
    join_geometry = nw.new_node(Nodes.JoinGeometry,
        input_kwargs={'Geometry': [transform_geometry, fillet_curve, extrude_mesh.outputs["Mesh"]]})
    
    group_output = nw.new_node(Nodes.GroupOutput, input_kwargs={'Geometry': join_geometry}, attrs={'is_active_output': True})

def apply(obj, selection=None, **kwargs):
    surface.add_geomod(obj, geometry_nodes, selection=selection, attributes=[])

\end{lstlisting}
\end{minipage}
\caption{Infinigen Blender Python Code for Creating a Table Model}
\label{fig:table_code}
\end{figure*}
\newpage

\end{document}